\documentclass[runningheads]{llncs}

\newif\ifcameraready
\camerareadytrue

\ifcameraready
  \usepackage{eccv}
\else
  \usepackage[review,year=2026,ID=10]{eccv}
\fi

\usepackage{eccvabbrv}
\usepackage{amsmath,amssymb}
\usepackage{graphicx}
\usepackage{booktabs}
\usepackage{array}
\usepackage{colortbl}
\usepackage{multirow}
\usepackage{pifont}
\usepackage{placeins}
\usepackage{tikz}
\usetikzlibrary{arrows.meta,positioning}
\usepackage[accsupp]{axessibility}

\ifcameraready
  \usepackage{hyperref}
\else
  \usepackage[pagebackref,breaklinks,colorlinks,citecolor=eccvblue]{hyperref}
\fi
\usepackage{orcidlink}

\definecolor{echoBlue}{HTML}{212F72}
\definecolor{echoTeal}{HTML}{3EC3B7}
\definecolor{echoAmber}{HTML}{3EC3B7} % legacy alias; yellow is intentionally unused
\definecolor{echoInk}{HTML}{000000}
\definecolor{echoGray}{HTML}{F1F3F5}
\newcommand{\cmark}{\textcolor{echoBlue}{\ding{51}}}
\newcommand{\xmark}{\textcolor{gray}{\ding{55}}}

\newcommand{\runin}[1]{\par\smallskip\noindent\textbf{#1}\ }
\newcolumntype{L}[1]{>{\raggedright\arraybackslash}p{#1}}

\hypersetup{
  pdftitle={Unlabeled Echoes: Pseudo-Labels and Genus-Aware Smoothing for Bat Call Recognition},
  pdfauthor={Frank Fundel and Alexandra Howard}
}

\begin{document}

\title{Unlabeled Echoes: Pseudo-Labels and Genus-Aware Smoothing for Bat Call Recognition}
\titlerunning{Unlabeled Echoes}

\ifcameraready
  \author{Frank Fundel\inst{1} \and Alexandra Howard\inst{2}}
  \authorrunning{F. Fundel and A. Howard}
  \institute{Ludwig-Maximilians-Universität München, Germany
  \and University of the Free State, Department of Zoology and Entomology, Qwaqwa Campus, South Africa}
\else
  \author{Anonymous CV4Ecology Submission}
  \authorrunning{Anonymous Authors}
  \institute{Anonymous institution}
\fi

\maketitle

\begin{abstract}
\setlength{\emergencystretch}{1em}
Passive acoustic monitoring produces far more bat recordings than experts can label.
We show that simple model-generated pseudo-labels turn this surplus into effective supervision.
We compare pseudo-labeling with other semi-supervised learning methods on an 18-species European corpus using only 10\% of its training labels, then transfer the strongest approaches to South African field audio containing nine bat taxa and a nuisance class.
Pseudo-labeling outperforms the other semi-supervised learning methods on every European measure, recovering up to 61.5\% of the gap to full supervision.
It transfers to field audio with gains of 10.69 points in species accuracy and 4.96 points in species macro-F1.
We also introduce genus-aware smoothing, which directs uncertain target mass toward congeneric species.
Combined with uniform smoothing, it reaches 79.16 species macro-F1, 4.73 points above hard targets.
Simple pseudo-labels are therefore highly effective at this ecological data scale, while genus-aware targets inject useful biological structure at no annotation cost.
\url{https://code4conservation.github.io/UnlabeledEchoes/}.
\end{abstract}

\section{Introduction}
\label{sec:intro}

Bats provide pest control, pollination, and seed dispersal, while their sensitivity to environmental change makes them valuable indicators of ecosystem condition~\cite{kunz2011ecosystem,jones2009carpe}.
Passive acoustic monitoring can sample entire nights of ultrasonic activity with little intervention.
Yet a recording archive is not an ecological result: species-level labels still require scarce expertise, and annotation effort grows with every deployed recorder~\cite{gibb2019emerging,stowell2022computational}.
Bat monitoring therefore presents a sharp asymmetry: collecting calls is increasingly cheap, but converting them into trustworthy supervision remains expensive.

Species recognition is intrinsically difficult.
Bat echolocation is shaped by behavioral and environmental tasks, creating substantial within-species variation and between-species overlap; automated systems consequently identify acoustic groups more reliably than many individual species~\cite{rydell2017prudence}.
Congeneric species are therefore biologically plausible alternatives, while devices, habitats, call phase, overlapping activity, and class imbalance further change the observed signal.
Automatic recognition has progressed from hand-crafted call descriptors and shallow classifiers~\cite{parsons2000acoustic,redgwell2009classification,walters2012continental} to convolutional and Transformer models operating on time--frequency representations~\cite{macaodha2018batdetective,chen2020tropical,fundel2023automatic}.
Modern models exploit labels more effectively but do not eliminate their cost; the largest resource is often the unannotated archive already being collected.

Semi-supervised methods are usually benchmarked on curated images or speech.
Their ranking remains unclear for moderate-scale, multi-label ecological audio.
Pseudo-labeling converts teacher predictions into task-specific targets and expands supervision \emph{across recordings}~\cite{lee2013pseudo}.
Genus-aware smoothing reserves uncertain target mass for congeneric species and structures supervision \emph{across classes}.

We test this claim in two complementary settings.
The European study masks 90\% of an 18-species training corpus; the South African study adds genuinely unlabeled field audio to labels for nine bat taxa and a nuisance output.
This changes geography, recording conditions, and source vocabulary, testing whether the conclusion survives outside label masking.
The classifier is based on the BioAcoustic Transformer (BAT) of Fundel et al.~\cite{fundel2023automatic}; fixing the compact backbone isolates how supervision is constructed.

Our contributions are:
\begin{itemize}
  \item We show that pseudo-labeling leads every measure in a controlled 10\%-label comparison, recovering up to 61.5\% of the gap to full supervision, and demonstrate that its species gains transfer to independent South African field audio.
  \item We introduce genus-aware smoothing, a target-level method that allocates uncertainty among congeneric species without additional annotations, and show in a controlled ablation that combining it with uniform smoothing improves species macro-F1 by 4.73 points over hard targets.
\end{itemize}

\section{Related Work}
\label{sec:related}

\runin{Bat and ecological acoustics.}
Early identification used manually measured call features and shallow classifiers~\cite{parsons2000acoustic,redgwell2009classification}; deep systems now reduce feature engineering and scale to larger archives~\cite{macaodha2018batdetective,chen2020tropical}.
Fundel et al. introduced BAT, the compact Transformer on which our classifier is based~\cite{fundel2023automatic}.
Most bat recognizers nevertheless remain supervised and inherit the biases of their annotations.
Bat2Web evaluates an SGAN with reduced label fractions, although its additional samples are generator-produced rather than unlabeled field calls~\cite{mahbub2024bat2web}.
Recent work studies cross-region bat recognition and self-supervised bat-song representations~\cite{macaodha2026general,deheerkloots2024exploring}, while reviews continue to identify domain shift and label quality as central bottlenecks~\cite{stowell2022computational}.

\runin{Bioacoustic benchmarks and pretrained models.}
BEANS, BirdSet, and iNatSounds provide cross-taxon, avian multi-label, and broad species-recognition benchmarks~\cite{hagiwara2023beans,rauch2025birdset,chasmai2024inatsounds}.
BirdNET, AVES, and the Perch family provide reusable predictions or embeddings, including transfer from birds to bats and other taxa; Perch~2.0 and MetaPerch further broaden pretraining~\cite{kahl2021birdnet,hagiwara2023aves,ghani2023perch,vanmerrienboer2025perch,chasmai2026metaperch}.
We instead fix an ultrasonic BAT backbone and vary how supervision is obtained.

\runin{Learning from unlabeled audio.}
SimCLR~\cite{chen2020simclr} and BYOL~\cite{grill2020byol} learn invariant representations from paired augmentations; audio-native alternatives include BYOL-A, SSAST, and AVES~\cite{niizumi2021byola,gong2022ssast,hagiwara2023aves}.
They optimize a proxy representation before fitting the species layer; self-training instead creates supervision directly in that output space.
Pseudo-labeling~\cite{lee2013pseudo}, Noisy Student~\cite{xie2020noisy}, and FixMatch~\cite{sohn2020fixmatch} differ mainly in teacher updates, perturbations, and consistency constraints.
CAP and ML-FixMatch+DA add class-aware positive/negative confidence control and distribution alignment, components absent from the retained implementation~\cite{xie2023cap,ihler2024multilabel}.
The families make different demands on augmentation, batches, and teacher reliability; we compare them at ecological scale and test whether their ranking transfers to field audio.
Active learning instead spends expert effort on model-selected recordings~\cite{qian2017active}; it is complementary to our fixed-label setting, which asks how much can be learned without requesting new annotations.

\runin{Taxonomic supervision.}
Uniform label smoothing regularizes classifiers by moving probability mass away from a hard target~\cite{szegedy2016inception}.
Contextual smoothing and hierarchy-aware losses instead encode phylogenetic or semantic proximity~\cite{trammell2019contextual,bertinetto2020mistakes}.
We bring this principle to multi-label bat-call recognition by concentrating smoothing mass on congeneric species, reflecting the documented ambiguity of species-level acoustic identification~\cite{rydell2017prudence}.
This target-level design requires neither an auxiliary hierarchy model nor further annotations, and remains compatible with ordinary uniform smoothing.

\section{Method}
\label{sec:method}

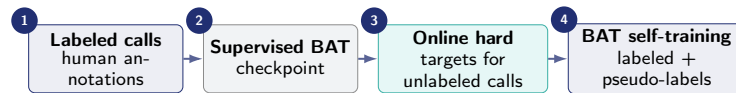
\begin{figure}[ht]
\centering
\begin{tikzpicture}[
  node distance=2.6mm,
  card/.style={draw=echoInk!25, line width=.35pt, rounded corners=2.4pt, fill=white, minimum height=9mm, text width=19mm, inner xsep=2pt, inner ysep=2pt, align=center, font=\sffamily\scriptsize},
  source/.style={card, draw=echoBlue, fill=echoBlue!8},
  model/.style={card, draw=echoInk!45, fill=echoGray},
  target/.style={card, draw=echoTeal!90!black, fill=echoTeal!13, text width=21mm},
  output/.style={card, draw=echoBlue, fill=echoBlue!8, text width=22mm},
  number/.style={circle, fill=echoBlue, text=white, minimum size=3.5mm, inner sep=0pt, font=\sffamily\tiny\bfseries},
  arrow/.style={-{Latex[length=1.8mm,width=1.2mm]}, line width=.75pt, draw=echoBlue!75}
]
\node[source] (human) {\textbf{Labeled calls}\\[-1pt]human annotations};
\node[model, right=of human] (teacher) {\textbf{Supervised BAT}\\checkpoint};
\node[target, right=of teacher] (pseudo) {\textbf{Online hard}\\targets for unlabeled calls};
\node[output, right=of pseudo] (student) {\textbf{BAT self-training}\\labeled + pseudo-labels};
\node[number, above left=-.8mm and -.8mm of human] {1};
\node[number, above left=-.8mm and -.8mm of teacher] {2};
\node[number, above left=-.8mm and -.8mm of pseudo] {3};
\node[number, above left=-.8mm and -.8mm of student] {4};
\draw[arrow] (human) -- (teacher);
\draw[arrow] (teacher) -- (pseudo);
\draw[arrow] (pseudo) -- (student);
\end{tikzpicture}
\caption{Retained pseudo-labeling protocol. The supervised checkpoint generates online hard targets while training continues on interleaved labeled and unlabeled batches.}
\label{fig:pipeline}
\end{figure}

\runin{BAT backbone.}
Our classifier is based on the BioAcoustic Transformer (BAT) introduced by Fundel et al.~\cite{fundel2023automatic}.
BAT divides each spectrogram sequence into 60 overlapping patches and maps every patch to a 64-dimensional token with five convolution--batch-normalization--ReLU blocks.
A learned classification token is appended before two self-attention layers with two heads and a feed-forward width of 32.
A linear layer and sigmoid activation produce the multi-label species probabilities.
For probability $p_c$ and target $y_c$ of class $c$, we use the asymmetric loss~\cite{benbaruch2021asymmetric}
\begin{equation}
\ell_{\mathrm{ASL}}(\mathbf p,\mathbf y)
 =\sum_c\left[-y_c(1-p_c)\log p_c-(1-y_c)p_c^2\log(1-p_c)\right].
\label{eq:asl}
\end{equation}
The first term penalizes missed positive classes, while the second penalizes false positives; the factors $(1-p_c)$ and $p_c^2$ reduce the influence of easy examples, especially abundant negatives.
Optimizer and schedule details are given in Appendix~\ref{app:implementation}.
The same compact backbone is retained across the comparison, isolating how supervision is constructed rather than changing model capacity.

\runin{Learning across recordings.}
A supervised BAT checkpoint is loaded and training continues on the same model; no separate frozen teacher is maintained (Fig.~\ref{fig:pipeline}).
At each self-training step, we set $\tilde{\mathbf y}$ to a one-hot vector: 1 for the highest-scoring class and 0 for every other class. Let $\hat{\mathbf p}$ be the current prediction. The unlabeled-data loss is scored with the ASL in Eq.~\ref{eq:asl}:
\begin{equation}
\mathcal{L}_U=\lambda\,\ell_{\mathrm{ASL}}(\hat{\mathbf p},\tilde{\mathbf y}).
\label{eq:pseudolabel}
\end{equation}
Here $\lambda$ controls the contribution of the unlabeled example; it is initialized at zero and increased to one during training. Targets are regenerated online, not stored, and no confidence threshold is used.
Human-labeled batches use the same ASL without the unlabeled-data weight.
Validation and test data are kept separate for evaluation.

\runin{Learning across classes.}
Uniform smoothing treats every wrong class alike.
Our \emph{genus-aware smoothing} instead places uncertain mass only on species that share the target genus (Fig.~\ref{fig:smoothing}a).
For a multi-hot label $\mathbf y$, let $m_k$ count how often class $k$ is a congeneric alternative and let $r_k$ normalize these counts:
\begin{equation}
\begin{aligned}
m_k&=\sum_{c:y_c=1}\mathbb{1}[k\in G(c)\setminus\{c\}],
&r_k&=\frac{m_k}{\max(\sum_jm_j,1)},\\
q_k&=(1-\beta)y_k+\beta\lVert\mathbf y\rVert_1r_k.
\end{aligned}
\label{eq:genus}
\end{equation}
Thus $\beta$ moves target mass to related species rather than distributing it uniformly.
The retained combined recipe uses genus smoothing with $\beta=.10$, followed by ordinary uniform smoothing with $\alpha=.02$.

\section{Experiments}
\label{sec:experiments}

\subsection{Task and Data}
\label{sec:data}

Given a time--frequency representation $x$, the model predicts probabilities $\hat{\mathbf p}\in[0,1]^C$ for a multi-hot target $\mathbf y\in\{0,1\}^C$.
We evaluate \emph{single-species} sequences with accuracy and unweighted per-species F1, and \emph{synthetic mixtures} made by combining one to three sequences and taking the union of their labels.
Mixtures use macro-F1 and test multi-label interference rather than natural acoustic overlap.

Skiba masks labels within a curated collection, enabling comparison with full supervision; UFS adds independent 256-kHz field audio whose 17-species source archive may contain taxa outside the model's ten outputs.
Both use the same spectrogram task and compact model but test different kinds of unlabeled data; exact counts are in Table~\ref{tab:data}.

\FloatBarrier
\begin{table}[ht]
\caption{Datasets used in the experiments. Splits are made at the recording level before sequence extraction with random seed 42.}
\label{tab:data}
\centering
\footnotesize
\renewcommand{\arraystretch}{.96}
\setlength{\tabcolsep}{2.2pt}
\resizebox{.99\linewidth}{!}{%
\begin{tabular}{@{}L{.9cm}L{2.15cm}L{2.35cm}L{2.75cm}L{2.95cm}@{}}
\toprule
Data & Source archive & Model outputs & Sequence split & Supervision used \\
\midrule
Skiba & $>$1,500 recordings; 29 species; 96~kHz~\cite{skiba2003europaische} & 18 species & 19,030 total: 11,186 train, 2,890 validation, 4,954 test & 1,119 labeled train (10\%); 10,067 train (90\%) treated as unlabeled; full reference uses all 11,186 \\
UFS & $>$40k recordings; 17 species; $\sim$15k without labels; 256~kHz & 9 bat taxa + Pesticide Spray & 15,863 total: 6,736 train, 3,857 validation, 5,270 test & 30,000 additional unlabeled field sequences \\
\bottomrule
\end{tabular}
}%
\end{table}

\subsection{Experimental Protocol}
\label{sec:protocol}

\runin{Input construction.}
Recordings are assigned to the training, validation, and test sets before sequences are created, so sequences from the same recording cannot appear in different sets.
Ultrasonic audio is time-dilated, high-pass filtered, downsampled, and converted to a decibel spectrogram.
Training mixtures combine one to three sequences and union their labels; labeled and unlabeled examples use the same acoustic construction.

\runin{Splits and supervision.}
The preparation script splits recordings within each metadata class into 60\% training, 20\% validation, and 20\% test sets (seed 42) before extracting and shuffling sequences.
The European comparison uses 10\% human labels and treats the other 90\% as unlabeled; the 100\% model is an upper reference.
BYOL, pseudo-labeling, and FixMatch are then evaluated on UFS.

\runin{Evaluation and reporting.}
Species F1 is the unweighted mean of the per-species F1 scores.
The BYOL row reports a four-run mean; remaining results are point estimates, and all gains below are absolute percentage points.

\runin{Comparison methods.}
Autoencoder pretraining reconstructs spectrogram patches, while BYOL aligns augmented views before classifier fitting~\cite{grill2020byol}.
Noisy Student uses a perturbed teacher and student~\cite{xie2020noisy}; FixMatch uses weak--strong consistency~\cite{sohn2020fixmatch}.
We also evaluate both cascade orders. These are comparison baselines; pseudo-labeling and genus-aware smoothing are the methods proposed here.

\FloatBarrier
\subsection{Results}
\label{sec:results}

\begin{table}[ht]
\caption{Skiba results (\%). Except for the gray full-supervision reference, all rows use 10\% labels; \cmark{} denotes unlabeled calls. Bold marks the best low-label result.}
\label{tab:europe}
\centering
\small
\setlength{\tabcolsep}{2.0pt}
\begin{tabular}{@{}lccccc@{}}
\toprule
Method & Labeled & Unlab. & Mixture F1 $\uparrow$ & Species acc. $\uparrow$ & Species F1 $\uparrow$ \\
\midrule
\rowcolor{echoGray} Full supervision & 100\% & \xmark & 72.46 & 84.00 & 80.33 \\
10\% baseline & 10\% & \xmark & 58.74 & 75.06 & 62.47 \\
\midrule
Autoencoder & 10\% & \cmark & 44.74 & 68.53 & 53.61 \\
BYOL & 10\% & \cmark & 52.68 & 69.54 & 57.69 \\
Noisy Student & 10\% & \cmark & 59.71 & 75.22 & 67.49 \\
FixMatch & 10\% & \cmark & 61.72 & 76.06 & 68.71 \\
Pseudo-label $\rightarrow$ FixMatch & 10\% & \cmark & 63.39 & 76.58 & 69.80 \\
FixMatch $\rightarrow$ Pseudo-label & 10\% & \cmark & 62.18 & 76.66 & 67.97 \\
\rowcolor{echoBlue!7} Pseudo-label & 10\% & \cmark & \textbf{65.12} & \textbf{80.48} & \textbf{73.45} \\
\bottomrule
\end{tabular}
\end{table}

\runin{Pseudo-labels recover up to 61.5\% of lost supervision.}
Pseudo-labeling leads all three European measures (Table~\ref{tab:europe}), adding 6.38 mixture-F1, 5.42 accuracy, and 10.98 species-F1 points over the matched baseline.
This closes \textbf{46.5\%, 60.6\%, and 61.5\%} of the respective gaps to full supervision.

\begin{table}[ht]
\caption{UFS results (\%). All methods use 6,736 labeled sequences; \cmark{} adds 30,000 unlabeled field sequences. Bold marks the best result.}
\label{tab:africa}
\centering
\small
\setlength{\tabcolsep}{7.0pt}
\begin{tabular}{@{}lcccc@{}}
\toprule
Method & Unlab. & Mixture F1 $\uparrow$ & Species acc. $\uparrow$ & Species F1 $\uparrow$ \\
\midrule
Supervised baseline & \xmark & 55.25 & 67.55 & 48.65 \\
BYOL & \cmark & 55.82 & 75.29 & 49.01 \\
FixMatch & \cmark & \textbf{57.98} & 76.99 & 49.04 \\
\rowcolor{echoBlue!7} Pseudo-label & \cmark & 56.69 & \textbf{78.24} & \textbf{53.61} \\
\bottomrule
\end{tabular}
\end{table}

\runin{Pseudo-labeling transfers to field audio.}
Every transferred learner exceeds the UFS baseline on all measures (Table~\ref{tab:africa}).
Pseudo-labeling gives the largest species gains (\textbf{+10.69 accuracy, +4.96 macro-F1}), while FixMatch leads on mixtures (+2.73 macro-F1).

\FloatBarrier
\begin{figure}[ht]
\centering
\includegraphics[width=\linewidth]{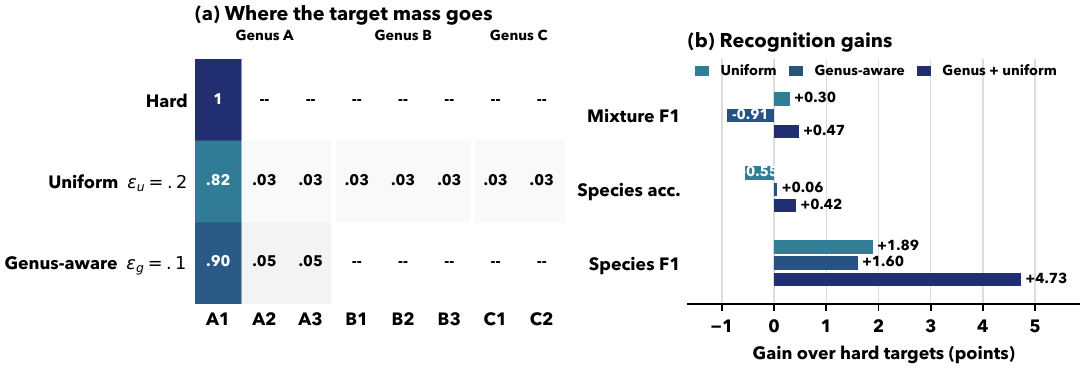}
\caption{Genus-aware smoothing and controlled compact-backbone results. (a) Target mass is restricted to related species. (b) Genus plus uniform smoothing gives the largest gains.}
\label{fig:smoothing}
\end{figure}

\runin{Controlled smoothing ablation.}
All four runs in Fig.~\ref{fig:smoothing} use the same data split, compact backbone, preprocessing, optimizer, and training schedule; only the construction of the training targets differs.
Genus-aware smoothing alone raises species macro-F1 from 74.43 to 76.03 (+1.60 points).
The combined recipe is strongest on every measure: 67.79 mixture F1 (+0.47), 81.51 species accuracy (+0.42), and \textbf{79.16 species F1 (+4.73)}; it also exceeds uniform smoothing by 2.84 species-F1 points.
Because accuracy weights sequences while macro-F1 weights species equally, the +0.42 versus +4.73 pattern indicates that smoothing chiefly benefits weaker classes.

\section{Discussion}
\label{sec:discussion}

Across both experiments, pseudo-labeling is the most consistent way to exploit additional audio with the fixed compact BAT backbone. It is the strongest low-label method on Skiba, recovering up to 61.5\% of the full-supervision gap, and improves the independent UFS field pool by 10.69 accuracy and 4.96 species-macro-F1 points. This transfer spans geography and recording conditions, while UFS remains limited to the model's known output classes.

Genus-aware smoothing complements pseudo-labeling by assigning uncertainty to biologically plausible congeneric alternatives. With the split, backbone, preprocessing, optimizer, and schedule fixed, its 4.73-point macro-F1 gain isolates target construction; the smaller 0.42-point accuracy gain indicates better performance on weaker species. These results use one European corpus, one field pool, and a fixed compact backbone, with hard online labels and a closed output vocabulary. Deployment should therefore pair predictions with confidence or open-set checks and expert review.

\enlargethispage{3\baselineskip}
\section{Conclusion}
\label{sec:conclusion}

This study shows that bat archives support semi-supervised recognition without exhaustive labeling. With 10\% of the European labels, pseudo-labeling recovers up to 61.5\% of the full-supervision gap and transfers to independent South African field audio. Genus-aware smoothing adds biologically structured uncertainty and improves species macro-F1 by 4.73 points without additional annotations. Together, these methods make unlabeled archives a practical resource for bat monitoring, with expert review for unfamiliar calls.

\clearpage
\bibliographystyle{splncs04}
\bibliography{main}

\clearpage
\appendix
\section{Implementation details}
\label{app:implementation}

The retained pseudo-label implementation interleaves unlabeled and labeled updates. A full pass over the human-labeled loader follows every ten unlabeled mini-batches. The pseudo-label weight is $\lambda_t=0$ before step 23, increases linearly as $(t-23)/10$ from steps 23 to 33, and remains 1 thereafter; the retained run starts at the beginning of this ramp. Human-labeled updates use unweighted asymmetric loss. The recording-level split and sequence-level masking are described in the main protocol and Table~\ref{tab:data}.

All main runs use SGD with a $5\times10^{-4}$ learning rate, cosine decay, and 46 epochs. The BYOL row is a four-run mean, whereas the remaining rows are point estimates.

\end{document}